\documentstyle[stwol]{article}
\input{psfig}
\def\Journal#1#2#3#4{{#1} {\bf #2}, #3 (#4)}

\def\NPB{{\em Nucl. Phys.} B}
\def\PLB{{\em Phys. Lett.}  B}

\def\PRD{{\em Phys. Rev.} D}
\def\ZPC{{\em Z. Phys.} C}

\def\be{\begin{equation}}
\def\ee{\end{equation}}
\def\bea{\begin{eqnarray}}
\def\eea{\end{eqnarray}}
\def\nn{\nonumber}
\newcommand{\spur}[1]{\not\! #1 \,}

\input psfig

\begin{document}

\title{ON THE THEORETICAL UNDERSTANDING \\
OF INCLUSIVE $\Lambda_b$ DECAYS }

\author{ PIETRO COLANGELO }

\address{Istituto Nazionale di Fisica Nucleare - Sezione di Bari -
via Amendola n.173, I-70126 Bari, Italy}


\twocolumn[\maketitle\abstracts{
I discuss a QCD sum rule determination of the  
four-quark operator matrix elements  contributing to the 
$\Lambda_b$ inclusive decay rates at ${\cal O}(m_b^{-3})$. 
The results suggest that $1/m_b^3$ corrections are not responsible 
of the observed difference between the lifetimes of $\Lambda_b$ and $B_d$.
}]

\section{Introduction}
The difference between the measured lifetimes of the
$\Lambda_b$ baryon and $B_d$ meson \cite{richman}:
\be 
\tau(\Lambda_b^0)/\tau(\bar {B^0}) = 0.78 \pm 0.04
\label{ratio}
\ee
represents an intriguing problem of the present-day heavy quark physics. 
Indeed, the $20 \%$ difference  contradicts the 
expectation that, at the scale of the $b$ quark mass, the spectator model
should describe rather accurately the decays of hadrons $H_Q$ containing one 
heavy quark. 

A calculation of the ratio
$\tau(\Lambda_b)/\tau(B_d)$ can be attempted using a field theoretical approach 
developed for the analysis of the inclusive weak 
decays of the hadrons $H_Q$ \cite{rev}. 
The method is based on the expansion in powers of
$m_Q^{-1}$, in the framework of the Wilson OPE.
The widths are expressed in terms of hadronic matrix elements 
of high dimensional quark and gluon operators;
since such matrix elements can be responsible of the
large difference between 
$\tau(\Lambda_b)$ and  $\tau(B_d)$, 
it is interesting to compute them
by nonperturbative methods such as QCD sum rules. 

The main aspects of the QCD analysis of the inclusive decays of hadrons $H_Q$ 
can be summarized considering 
the transition operator $\hat T(Q \to X_f \to Q)$ 
\cite{rev}:
\begin{equation}
\hat T=i \int d^4 x \; T[{\cal L}_W(x){\cal L}_W^\dagger(0)] 
\label{t} 
\end{equation}
\noindent 
which describes an amplitude with the heavy quark $Q$
having the  same momentum in the initial and final state.
${\cal L}_W$ is the effective weak Lagrangian governing the decay
$Q \to X_f$.
The inclusive width $H_Q\to X_f$ can be obtained from
\begin{equation}
\Gamma(H_Q \to X_f)={ 2 \; Im~<H_Q|\hat T|H_Q> \over 2\; M_{H_Q}} \hskip 3 pt.
\label{width} 
\end{equation}
\noindent 
The large energy release in the heavy quark decay permits 
an expansion of 
$\hat T$ in terms of local operators ${\cal O}_i$:
\begin{equation}
\hat T= \sum_i C_i {\cal O}_i \;\;\; \label{ope}
\end{equation}
with ${\cal O}_i$ ordered according to the dimension, and the 
coefficients $C_i$ containing appropriate inverse powers of the heavy quark 
mass $m_Q$.
The lowest dimension operator in (\ref{ope}) is ${\cal O}_3=\bar Q Q$.
The next gauge and Lorentz invariant operator is the $D=5$ chromomagnetic 
operator
${\cal O}_G= {\bar Q} {g \over 2}  \sigma_{\mu \nu} G^{\mu \nu} Q$, whose 
hadronic matrix element
\begin{equation}
\mu_G^2(H_Q)={ <H_Q| {\bar Q} {g \over 2} \sigma_{\mu \nu} G^{\mu \nu} Q|H_Q> 
\over 2 M_{H_Q} }  \label{chro} 
\end{equation}
\noindent 
measures the coupling of the heavy quark spin to the spin of the light degrees 
of freedom in $H_Q$, and therefore 
is responsible of the mass splitting between
hadrons belonging to the same $s_\ell$ multiplet
($s_\ell$ is the total angular momentum of the light degrees of freedom in 
$H_Q$). In the case of $b$-flavoured hadrons
this mass difference has been measured,
both for mesons ($M_{B^*} - M_B = 45.7\pm \; 0.4\; MeV$) and 
$\Sigma_b$ baryons \cite{delphi} 
($M_{\Sigma^*_b} - M_{\Sigma_b} = 56 \pm 16 \; MeV$).

The matrix element of $\bar Q Q$ over $H_Q$ can be obtained using the 
following expansion stemming from the heavy 
quark equation of motion:
\begin{equation}
\bar Q Q = \bar Q \gamma^0 Q + { {\cal O}_G \over 2 m_Q^2} -
{ {\cal O}_\pi \over 2 m_Q^2} + \; O(m_Q^{-3}) \; \; \; ;
\end{equation}
${\cal O}_\pi$ is the kinetic energy operator 
${\cal O}_\pi={\bar Q} (i {\vec D})^2 Q$ whose matrix element
\begin{equation}
\mu_\pi^2(H_Q)={ <H_Q| {\bar Q} (i {\vec D})^2 Q |H_Q> \over 2 M_{H_Q} } 
\label{kin} 
\end{equation}
measures the average squared momentum of the heavy quark inside $H_Q$.

At the order  ${\cal O}(m_Q^{-3})$ 
four-quark operators appear in (\ref{ope}):
\be
{\cal O}_6^q = {\bar Q} \Gamma q \; {\bar q } \Gamma Q \label{4q}
\ee
with $\Gamma$ an appropriate combination of Dirac and color matrices.

The resulting expression for the width
$\Gamma(H_Q \to X_f)$ reads:
\be
\Gamma(H_Q \to X_f)=\Gamma_0^f \; \Big[ A_0^f 
+{A_2^f \over m_Q^2} +{A_3^f \over m_Q^3} +
...\Big] \;\;.\label{expan}
\ee
$A_i^f$ and $\Gamma_0^f$ depend on the 
final state $X_f$; $A_i^f$   
include perturbative short-distance coefficients and  
nonperturbative hadronic matrix elements incorporating the
long range dynamics.
The leading term  in (\ref{expan}) corresponds to
the partonic prediction 
$\Gamma^{part}(H_Q \to X_f)=\Gamma_0^f A_0^f$, with $A_0^f=1+ c^f {\alpha_s 
/ \pi}+ O(\alpha_s^2)$ and $\Gamma_0^f \propto m_Q^5$; 
differences among the widths of the  
hadrons $H_Q$  emerge at the next to leading order in $1/m_Q$, and are related 
to the different value of the matrix elements of the operators ${\cal O}_i$
of dimension larger than three.

It is important to notice the absence of 
the first order term $m_Q^{-1}$ in (\ref{expan}) \cite{chay}.

The $D=5$ operators ${\cal O}_G$ and ${\cal O}_\pi$ 
are SU(3) singlets; on the contrary,
the $D=6$ operators in (\ref{4q})
give different contributions when averaged over
hadrons belonging to the same $SU(3)$ light flavour multiplet, and therefore
they are responsible of the different lifetimes  
of, e.g., $B^-$ and $B_s$, $\Lambda_b$ and $\Xi_b$.
The spectator flavour dependence is related to the mechanisms of weak 
scattering and Pauli interference \cite{rev}, both suppressed by the factor 
$m_Q^{-3}$ with respect to the parton decay rate.

As for differences between mesons and baryons, they could
already arise at ${\cal O}(m_Q^{-2})$, due both to the chromomagnetic
contribution and to the kinetic energy term in (\ref{width}). 
In particular, the kinetic energy term is responsible of the 
difference for systems where the chromomagnetic contribution 
vanishes, namely in the case of $\Lambda_b$ and $\Xi_b$ having 
the light degrees of freedom in $S-$ wave. 
However, the results of a
calculation of $\mu^2_\pi$ for mesons \cite{braun}
and baryons \cite{noi} support the conjecture \cite{wise}
that the kinetic energy operator has the same matrix element when computed on 
such hadronic systems. \cite{ekin} 
The approximate equality of the kinetic energy operator on $B_d$ 
and $\Lambda_b$ can also be inferred by  using mass relations
\cite{bigi2}:
$\mu^2_\pi(\Lambda_b)- \mu^2_\pi(B_d) \simeq
0.002\pm 0.024~GeV^2$.
Then,
differences between meson and baryon lifetimes should  
occur at the $m_Q^{-3}$ level, thus involving the four-quark operators
in (\ref{4q}). They can be classified as follows  \cite{neub}:
\begin{eqnarray}
O^q_{V-A}&=&{\bar Q}_L \gamma_\mu q_L \; {\bar q}_L \gamma_\mu Q_L \nonumber \\
O^q_{S-P}&=&{\bar Q}_R q_L \; {\bar q}_L  Q_R \nonumber \\
T^q_{V-A}&=&{\bar Q}_L \gamma_\mu {\lambda^a \over 2} q_L \;
{\bar q}_L \gamma_\mu {\lambda^a \over 2} Q_L \nonumber \\
T^q_{S-P}&=&{\bar Q}_R {\lambda^a \over 2} q_L \; {\bar q}_L  
{\lambda^a \over 2} Q_R \label{4q1}
\end{eqnarray}
($q_{R,L}= {1 \pm \gamma_5 \over 2} q$, $\lambda_a$ Gell-Mann matrices).

For mesons,  the matrix elements of the operators in (\ref{4q1})
can be computed by vacuum saturation approximation:
\begin{eqnarray}
<B_q | O^q_{V-A} | B_q>_{VSA} &=& {f_{B_q}^2 M_{B_q}^2 \over 4}\\
<B_q | T^q_{V-A} | B_q>_{VSA} &=& 0 \;\;\; ,\label{vsa}
\end{eqnarray}
etc. 
Such an approximation cannot be employed for baryons, where
a direct calculation is required.

For $\Lambda_b$, a simplification can be
obtained \cite{neub} introducing 
\begin{equation}
\tilde{\cal O}^q_{V-A} = {\bar Q}^i_L \gamma_\mu Q^i_L \; 
{\bar q}^j_L \gamma^\mu q^j_L \label{otilde}
\end{equation}
and
\begin{equation}
\tilde{\cal O}^q_{S-P} = {\bar Q}^i_L  q^j_R \; 
{\bar q}^j_L Q^i_R 
\end{equation}
($i$ and $j$ color indices): 
the $\Lambda_b$ matrix elements of the  operators in (\ref{4q1})
can be expressed in terms of 
$<\Lambda_b | \tilde{\cal O}^q_{V-A}|\Lambda_b>$ and
$<\Lambda_b | {O}^q_{V-A}|\Lambda_b>$, modulo $1/m_Q$ corrections 
contributing to subleading terms in the expression for the inclusive widths.

Parametrizing the matrix elements
\begin{equation}
\langle \tilde {\cal O}^q_{V-A}\rangle_{\Lambda_b} =
{<\Lambda_b | \tilde {\cal O}^q_{V-A} |\Lambda_b> \over 2 M_{\Lambda_b} } =
{f_B^2 M_B \over 48} r  \label{par}
\end{equation}
\begin{equation}
{<\Lambda_b | {O}^q_{V-A} |\Lambda_b>  } = - {\tilde B} \;\;
{<\Lambda_b | \tilde {\cal O}^q_{V-A} |\Lambda_b>  } \;, \label{btilde}
\end{equation}
one has that, for $f_B=200 \; MeV$ and $r=1$,
 eq.(\ref{par}) corresponds to 
$\langle \tilde {\cal O}^q_{V-A}\rangle_{\Lambda_b} = 4.4 \times 10^{-3} \;
GeV^3$.
Quark models predict 
$\langle \tilde {\cal O}^q_{V-A}\rangle_{\Lambda_c} \simeq 0.75 - 2.5 
\times 10^{-3} 
\; GeV^3$, corresponding to
$r \simeq 0.2 - 0.6$ \cite{bilic};
larger values 
 can be obtained \cite{rosner}
using the mass splitting $\Sigma_b^* - \Sigma_b$ and
$\Sigma_c^* - \Sigma_c$:
$r \simeq 0.9 - 1.8$.
In the valence quark approximation $\tilde B=1$.

A value of $r$: $r\simeq 4\;-\;5$ 
would explain the difference between $\tau(\Lambda_b)$ and $\tau(B_d)$
\cite{neub}. A calculation by QCD sum rules, however, seems to exclude this 
possibility.

\section{$\langle{\tilde {\cal O}^q_{V-A}}\rangle_{\Lambda_b}$ 
from QCD sum rules}
An estimate of 
$\langle{\tilde {\cal O}^q_{V-A}}\rangle_{\Lambda_b}$ 
can be obtained by 
QCD sum rules, in HQET, analyzing the three-point correlator
\bea
&&\Pi_{CD}(\omega, \omega^\prime )=(1+ {\spur v})_{CD} 
\Pi(\omega, \omega^\prime)\nn\\
&=& i^2 \int dx dy <0| T [J_{C}(x) {\tilde {\cal O}^q_{V-A}}(0)
{\bar J}_{D}(y)] |0> \nn\\
&\times& e^{i \omega v \cdot x  - i \omega^\prime v \cdot y}
\label{threep}
\eea
of the baryonic currents $J (x)$ and ${\bar J}(y)$ 
and of the operator $\tilde {\cal O}^q_{V-A}$ in (\ref{otilde}).
The variables $\omega$, $\omega^\prime$ are related to the residual momentum
of the currents:
$p^\mu = m_b v^\mu + \omega v^\mu$.
Choosing $J$ with a non-vanishing
projection on the $\Lambda_b$ state
\begin{equation}
<0| J_C | \Lambda_b(v)> = f_{\Lambda_b} (\psi_v)_C \;\label{fl}
\end{equation}
($\psi_v$ is a spinor for a $\Lambda_b$ of four-velocity $v$),
the parameter 
$f_{\Lambda_b}$ representing the coupling of $J$ to  
$\Lambda_b$, the matrix element  
$\langle{\tilde {\cal O}^q_{V-A}}\rangle_{\Lambda_b}$ 
can be obtained by saturating the correlator (\ref{threep}) 
with baryonic states, 
and considering the double pole contribution  in the variables $\omega$ 
and  $\omega^\prime$:
\begin{equation}
\Pi^{h}(\omega, \omega^\prime) = 
\langle{\tilde {\cal O}^q_{V-A}}\rangle_{\Lambda_b}
{f^2_{\Lambda_b} \over 2} {1 \over (\Delta_{\Lambda_b} - \omega)
(\Delta_{\Lambda_b} - \omega^\prime)} 
\label{phad}
\end{equation}
at the value  $\omega=\omega^\prime=\Delta_{\Lambda_b}$. 
The mass parameter $\Delta_{\Lambda_b}$ represents the binding energy of 
the light $\Lambda_b$ degrees of freedom
in the static color field generated by the $b-$quark:
$M_{\Lambda_b}= m_b + \Delta_{\Lambda_b}$; it
must also be derived by QCD sum rules.

A possible interpolating field for $\Lambda_b$ 
\cite{shuryak} reads:
\begin{equation}
J_C(x)= \epsilon^{i j k} ( q^{T i}(x) \Gamma \tau q^j(x))
(h_v^k)_C(x) \label{curr}
\end{equation}
where $T$ means transpose,  $i, j$ and $k$ are color indices, and
 $C$ is the Dirac index of 
the effective heavy quark field $h_v(x)$;
$\tau$ is a light flavour matrix corresponding to zero isospin.
In the $m_b \to \infty$ limit, 
the light diquark in the $\Lambda_b$  is in a relative $0^+$ spin-parity state; 
therefore one can use in (\ref{curr}): 
\be
\Gamma= C \gamma_5 ( 1 + b {\spur v} \;) \label{curr1}
\ee
$C$ being the charge conjugation matrix and $b$ a parameter. 
Arguments can be found 
in favour of the choice $b=1$ in (\ref{curr1}) \cite{noi}. 

The coupling $f_{\Lambda_b}$ can be computed from
\cite{noi,grozin}
\bea
&&H_{CD}(\omega) =
(1+ {\spur v})_{CD} H(\omega) \nn \\
&=& 
i \int dx  <0| T [J_{C}(x) {\bar J}_{D}(0)] |0> e^{i \omega v \cdot x} \;.
\hfill \label{twop}
\eea
In the Euclidean region 
the correlation functions (\ref{threep})
and (\ref{twop}) can be computed in QCD, in terms of a 
perturbative contribution and vacuum condensates:
\begin{equation}
\Pi^{OPE}(\omega, \omega^\prime)= \int d\sigma d\sigma^\prime 
{\rho_\Pi(\sigma, \sigma^\prime) \over(  \sigma - \omega ) (\sigma^\prime
- \omega^\prime)} \label{pope}
\end{equation}
\begin{equation}
H^{OPE}(\omega)= \int d\sigma {\rho_H(\sigma) \over(\sigma - \omega) } \;,
\end{equation}
with the 
spectral functions given by :
\bea
\rho_\Pi(\sigma, \sigma')&=&\rho^{(pert)}_\Pi(\sigma,\sigma') 
+ \rho^{(D=3)}_\Pi(\sigma,\sigma') <{\bar q} q> \nonumber \nn \\
&+& \rho^{(D=4)}_\Pi(\sigma,\sigma') < {\alpha_s \over \pi} G^2> \nn \\
&+& \rho^{(D=5)}_\Pi(\sigma,\sigma') <{\bar q}g \sigma G  q> \nn \\
&+& \rho^{(D=6)}_\Pi(\sigma,\sigma') (<{\bar q} q>)^2 + \dots \;\; 
\label{specf}
\eea
and a a similar expression for $\rho_H(\sigma)$.

The various terms in $\rho_\Pi(\sigma, \sigma^\prime)$  and
$\rho_H(\sigma)$ 
can be found in \cite{noi,defazio}.

A sum rule for $\langle \tilde {\cal O}^q_{V-A}\rangle_{\Lambda_b}$ 
can be derived by equating the hadronic and 
the OPE representations of the correlator (\ref{threep}),
and modeling 
the contribution of the higher resonances and of the continuum in 
$\Pi^{h}$ as the QCD term  outside the region 
$0 \le \omega \le \omega_c$,
$0 \le \omega^\prime \le \omega_c$, $\omega_c$ being an effective threshold.
After a double Borel transform one gets:  
\bea
&&{f^2_{\Lambda_b} \over 2} (1 + b)^2 e^{- {\Delta_{\Lambda_b} \over E}} 
\langle \tilde {\cal O}^q_{V-A}\rangle_{\Lambda_b} \\
&=&
\int_0^{\omega_c} \int_0^{\omega_c} d \sigma d \sigma^\prime
e^{-{ \sigma + \sigma^\prime \over 2 E} } 
\rho_\Pi(\sigma, \sigma^\prime)
\;\;\; \label{sr}
\eea
where $E$ is a Borel parameter.
The threshold $\omega_c$ can be fixed  
in the QCD sum rule determination of $f_{\Lambda_b}$ and 
$\Delta_{\Lambda_b}$ 
\cite{noi}: $\omega_c=1.1 - 1.3 \; GeV$; in correspondence one obtains 
$f_{\Lambda_b}=(2.9 \pm 0.5) \times 10^{-2} \; GeV^3$ 
and $\Delta_{\Lambda_b}=0.9 \pm 0.1 \; GeV$.

\begin{figure}[ht]
\center
\mbox{\psfig{figure=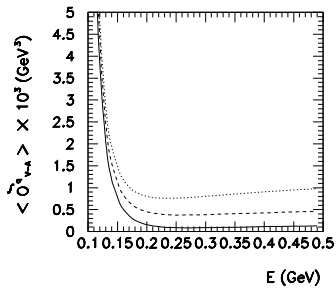,height=1.5in}}
\caption{Sum rule for the matrix element 
$\langle{\tilde {\cal O}^q_{V-A}}\rangle_{\Lambda_b}$}
\label{fig:sumr}
\end{figure}

The resulting sum rule is depicted in fig.1.
A stability window is observed for $E>0.2 \; GeV$.
In the duality region $E\simeq 0.2 - 0.3 \; GeV$
(the same region considered in the QCD sum rule
analysis of $f_{\Lambda_b}$ and $\mu^2_\pi(\Lambda_b)$ \cite{noi})
we find:
\begin{equation}
\langle\tilde {\cal O}^q_{V-A}\rangle_{\Lambda_b} \simeq (0.4 - 1.20) \times 
10^{-3} \; GeV^3 \;\;\; , \label{res}
\end{equation}
which corresponds to $r\simeq 0.1~-~0.3$.
This result is confirmed by an analysis based on the assumption of
local quark-hadron duality,
that amounts to calculate the matrix elements
of  $\langle\tilde {\cal O}^q_{V-A}\rangle_{\Lambda_b}$ and $f_{\Lambda_b}$ 
by free quark 
states produced and annihilated by the baryonic currents in (\ref{threep}) and 
(\ref{twop}), and then averaging on a duality interval in $\omega, 
\omega'$ \cite{defazio}.

Considering, finally, the parameter $\tilde B$ in eq.(\ref{btilde}),
one gets $\tilde B=1$ since, 
in this computational scheme, only valence quark processes 
are taken into account.

\section{Conclusions}

Within the uncertainties of the method, QCD sum rules predict 
small values for the matrix elements 
$\langle \tilde {\cal O}^q_{V-A} \rangle_{\Lambda_b}$ and
$\langle  {\cal O}^q_{V-A} \rangle_{\Lambda_b}$, comparable with the outcome of 
constituent quark models. 
The conclusion is 
that the inclusion of $1/m_Q^3$ terms in the 
expression of the inclusive widths does not solve the puzzle represented by the 
difference between 
$\tau(\Lambda_b)$ and $\tau(B_d)$:
using the formulae in \cite{neub} for the lifetime ratio, 
the value in  (\ref{res}) together with $\tilde B=1$ gives:
\begin{equation}
\tau(\Lambda_b)/\tau(B_d) \ge 0.94 \; \; \;.
\end{equation}

It seems unlikely that order $m_Q^{-4}$ terms can solve 
the problem. We must conclude that,  if the measurement of 
$\tau(\Lambda_b)$ and $\tau(B_d)$ will be confirmed,
a reanalysis of the problem will be required.
In particular, it has been proposed to consider the possibility of the
failure of the 
assumption, made in the calculation of the lifetimes, of local quark-hadron 
duality in nonleptonic inclusive decays
\cite{falk,alt,mart}. 
Meanwhile, it is interesting that new data 
are now available for other $b-$flavoured hadrons, e.g.
$\Xi_b$ \cite{richman}, although with errors too large to perform a meaningful
comparison with $\Lambda_b$. Such new information will be of great
importance for 
the full understanding of the problem of the beauty hadron lifetimes.

\section*{Acknowledgments}
I would like to thank F.De Fazio, C.A.Dominguez, G.Nardulli and
N.Paver for the pleasant collaboration on the topics discussed in this talk.


\section*{References}
\begin{thebibliography}{99}

\bibitem{richman}
J.Richman, these Proceedings.

\bibitem{rev}
B.Blok and M.Shifman in {\em Proceedings of the Third Workshop
on the Tau-Charm Factory}, eds. J.Kirkby and R.Kirkby, (Ed. Frontieres, 1994)
p.269, and references therein.

\bibitem{delphi}
P.Abreu {\it et al.}, DELPHI Collaboration,  
preprint DELPHI 95-107 PHYS 542 (1995).

\bibitem{chay}
J.Chay {\it et al.}, \Journal{\PLB}{247}{399}{1990};
I.Bigi {\it et al.}, \Journal{\PLB}{293}{430}{1992}; {\bf 297},477 (1993) (E).

\bibitem{braun}
P.Ball and V.M.Braun, \Journal{\PRD}{49}{2472}{1994}.

\bibitem{noi}
P.Colangelo, C.A.Dominguez, G.Nardulli and N.Paver,
\Journal{\PRD}{54}{4622}{1996}.

\bibitem{wise}  
A.V.Manohar and M.B.Wise, \Journal{\PRD}{49}{1310}{1994}.

\bibitem{ekin}
The value of $\mu^2_\pi$ is quite controversial: for a discussion see
M.Neubert, hep-ph/9610471 and references therein.  
 
\bibitem{bigi2} 
I.Bigi, hep-ph/9508408.

\bibitem{neub}
M.Neubert and C.T.Sachrajda, hep-ph/ 9603202.

\bibitem{bilic}
B.Guberina {\it et al.}, \Journal{\ZPC}{33}{297}{1986}.

\bibitem{rosner}
J.L.Rosner, \Journal{\PLB}{379}{267}{1996}.

\bibitem{shuryak}
E.V.Shuryak, \Journal{\NPB}{198}{83}{1982}.

\bibitem{grozin} A.G.Grozin {\it et al.}, \Journal{\PLB}{285}{254}{1992};
{\bf 291}, 441 (1992);
E.Bagan {\it et al.}, \Journal{\PLB}{278}{367}{1992}; {\bf 287}, 176 (1992); 
{\bf 301}, 243 (1993);
Y.B.Dai {\it et al.}, \Journal{\PLB}{371}{99}{1996};
S.Groote {\it et al.}, \Journal{\PRD}{54}{3477}{1996}; hep-ph/9609469.

\bibitem{defazio}
P.Colangelo and F.De~Fazio, \Journal{\PLB}{387}{371}{1996}.

\bibitem{falk}
A.Falk, M.B.Wise and I.Dunietz, \Journal{\PRD}{51}{1183}{1995}.

\bibitem{alt}
G.Altarelli {\it et al.}, \Journal{\PLB}{382}{409}{1996}.

\bibitem{mart}
G.Martinelli, these Proceedings.

\end{thebibliography}
\end{document}